\documentclass{article} 
\PassOptionsToPackage{table}{xcolor}
\usepackage[final]{colm2026_conference}

\usepackage[T1]{fontenc}
\usepackage{microtype}
\usepackage{graphicx}
\usepackage{wrapfig}
\usepackage[most]{tcolorbox}
\usepackage{subcaption}
\usepackage{booktabs}
\usepackage{multirow}
\usepackage{enumitem}
\usepackage{xcolor}
\usepackage{pifont}
\usepackage{stfloats}
\usepackage{array}
\usepackage{makecell}
\usepackage{tikz}
\usepackage{pgf-pie}
\newcommand{\cmark}{\textcolor{mygreen}{\ding{51}}}
\newcommand{\xmark}{\textcolor{myred}{\ding{55}}}

\usepackage{hyperref}
\definecolor{prettyred}{HTML}{FC4C4E}
\definecolor{mygreen}{RGB}{25, 110, 2}
\definecolor{myred}{RGB}{173,3,3}

\newcommand{\pmark}{%
  \tikz[baseline=-0.6ex]{
    \draw (0,0) circle (0.9ex);
    \begin{scope}
      \clip (-0.9ex,-0.9ex) rectangle (0,0.9ex); 
      \fill (0,0) circle (0.9ex);
    \end{scope}
  }%
}

\usepackage{pgfplots}
\pgfplotsset{compat=1.18}

\usepackage{macros}

\usepackage{amsmath}
\usepackage{amssymb}
\usepackage{mathtools}
\usepackage{amsthm}
\usepackage{listings}

\usepackage[capitalize,noabbrev]{cleveref}

\usepackage{url}

\usepackage{lineno}

\definecolor{darkblue}{rgb}{0, 0, 0.5}
\hypersetup{colorlinks=true, citecolor=darkblue, linkcolor=darkblue, urlcolor=darkblue}

\usepackage[textsize=tiny]{todonotes}

\title{\toolname: Repository-Scale Formal Verification Benchmarks for Lean}

\newcommand{\utaustinmark}{\ensuremath{\spadesuit}}
\newcommand{\nyumark}{\ensuremath{\diamondsuit}}
\newcommand{\ualbertamark}{\ensuremath{\clubsuit}}
\newcommand{\amiimark}{\ensuremath{\heartsuit}}
\author{%
  \bfseries Yutong Xin\textsuperscript{\utaustinmark}\thanks{Equal contribution.}
  \quad Jocelyn Qiaochu Chen\textsuperscript{\nyumark\,\ualbertamark\,\amiimark}\footnotemark[1]
  \quad Greg Durrett\textsuperscript{\nyumark}
  \quad I\c{s}il Dillig\textsuperscript{\utaustinmark}\\[0.35em]
  \normalfont
  \textsuperscript{\utaustinmark}The University of Texas at Austin
  \quad \textsuperscript{\nyumark}New York University\\
  \textsuperscript{\ualbertamark}University of Alberta
  \quad \textsuperscript{\amiimark}Alberta Machine Intelligence Institute (Amii)\\[0.25em]
  \texttt{maxryeery@utexas.edu}
  \quad \texttt{jocelyn.chen@ualberta.ca}
}

\begin{document}

\ifcolmsubmission
\linenumbers
\fi

\maketitle

\begin{abstract}
Large language models have achieved striking results in interactive theorem proving, particularly in Lean. However, most benchmarks for LLM-based proof automation are drawn from mathematics in the Mathlib ecosystem, whereas proofs in software verification are developed inside definition-rich codebases with substantial project-specific libraries. We introduce \toolname, a benchmark of 500 Lean~4 proof obligations drawn from open-source formal-methods developments and packaged to preserve realistic repository context and cross-file dependencies. Our evaluation of frontier LLMs and specialized provers yields three observations. First, provers tuned for Mathlib-style mathematics transfer poorly to this repository-centric setting. Second, success is strongly correlated with transitive repository dependence: tasks whose proofs draw on large, multi-hop dependency closures are less likely to be solved.  Third, providing curated context restricted to a proof's dependency closure improves performance relative to exposing the full repository, but nevertheless leaves substantial room for improvement. 
Our benchmark and evaluation suite are released at \href{https://github.com/utopia-group/VeriSoftBench}{https://github.com/utopia-group/VeriSoftBench}.
\end{abstract}

\section{Introduction}
Interactive theorem provers (ITPs) such as Lean~\citep{lean}, Rocq/Coq~\citep{rocq}, and Isabelle~\citep{isabelle} provide machine-checked guarantees, but this rigor comes with a well-known cost: even routine mathematical arguments require low-level, often tedious formalization, making mechanized proof development extremely labor-intensive. To reduce this burden, decades of work have pursued automation within ITP workflows, for example, using domain-specific tactic libraries~\citep{ssreflect, lean-egg} or integration with external solvers~\citep{lean-smt,hammer, thor}. More recently, large language models (LLMs) have enabled a new class of \emph{neural} proof automation methods~\citep{goedelv2, deepseek-prover-v2, hilbert}, and high-quality benchmarks have become crucial for measuring progress in the proof settings where automation is ultimately needed.

\begin{figure}[ht!]
    \centering
    \includegraphics[width=\linewidth]{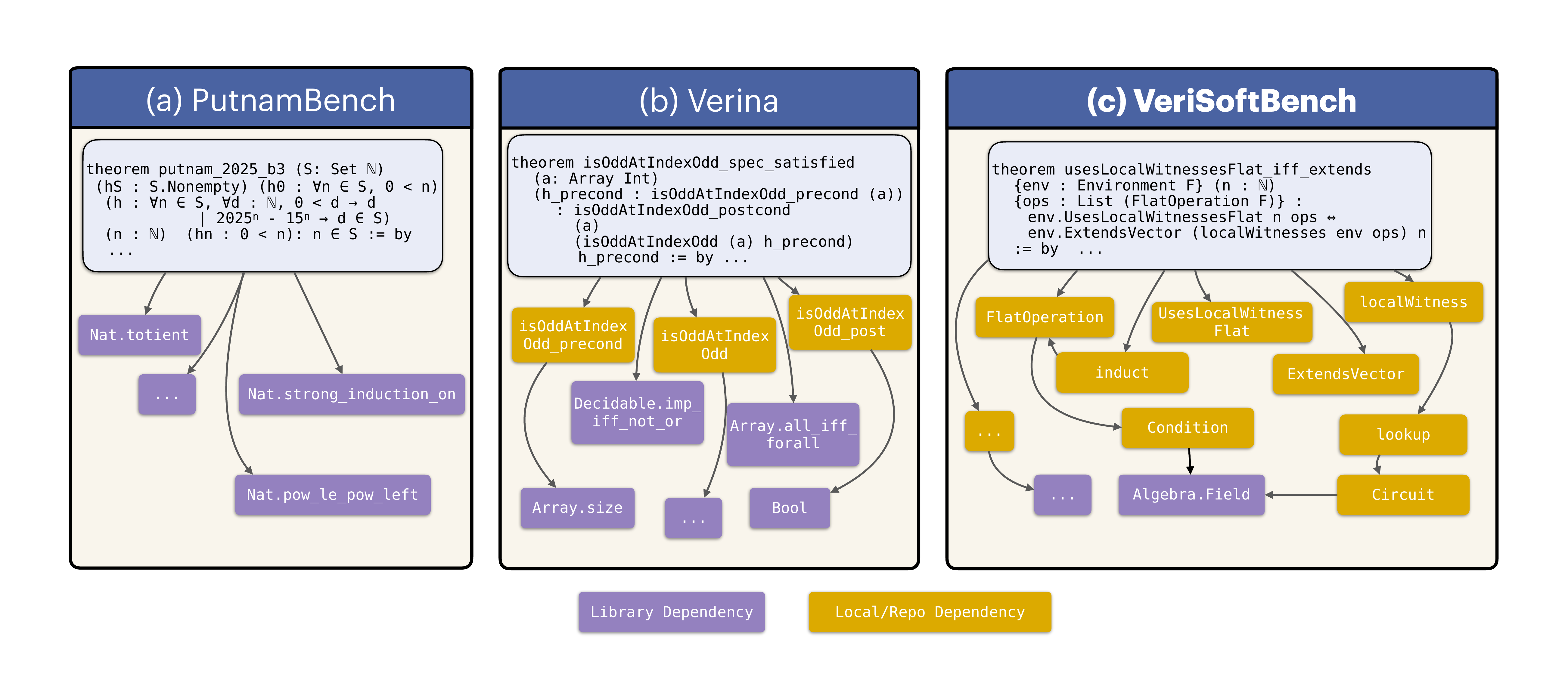}
    \vspace{-1.5em}
    \caption{Contextual dependencies comparison between (a) mathematical benchmark proofs, (b) lightweight verification tasks, and (c) repository-scale verification. PutnamBench proofs rely almost entirely on library (Mathlib) dependencies (purple). Verina introduces a small number of project-specific definitions (yellow) but remains largely library-driven. In contrast, proofs from a real-world ZK circuit DSL repository~\citep{clean} depend heavily on interconnected local abstractions, forming a dense repository-level dependency graph. Purple denotes library dependencies; yellow denotes local/repository dependencies.}
    \label{fig:mot}
\end{figure}

Most widely used benchmarks for LLM-based proof automation are drawn from \emph{mathematics} and operate in a shared ecosystem centered around Mathlib. Suites such as MiniF2F~\citep{minif2f}, ProofNet~\citep{proofNet}, LeanDojo~\citep{leanDojo}, and PutnamBench~\citep{putnamBench} have been instrumental in driving rapid improvements. In these settings, proof goals are typically stated over well-known mathematical structures that are already formalized in shared libraries, so the prover can rely heavily on pre-existing definitions and lemmas. Formal verification, by contrast, presents a structurally different landscape: projects that verify compilers, protocol implementations, or language runtimes define bespoke syntax, semantics, and domain abstractions with no counterpart in shared libraries. Consequently, proof obligations depend on deep chains of project-specific definitions distributed across multiple files (Figure~\ref{fig:mot}). This raises a key question: \emph{How well do current provers generalize when proofs are grounded in large, project-specific libraries rather than a shared ecosystem?}

This question is particularly pressing given how neural provers have evolved. Earlier tactic-prediction systems~\citep{astactic,tactok,htps, gpt-f} interact with the proof assistant step by step, with the kernel managing the environment visible at each goal. Recent LLM-based provers~\citep{goedelv2, deepseek-prover-v2} instead consume a prompt containing the proof obligation together with repository context and generate a complete proof. In this paradigm, \emph{what context to provide}---and how the prover reasons over it---becomes a central challenge. However, Lean~4, where much of recent prover development is concentrated, lacks a benchmark that targets this challenge in the verification setting.

This paper introduces \toolname, a benchmark of 500 Lean~4 proof obligations drawn from open-source formal-methods developments and packaged to preserve cross-file dependency structure. By preserving the original project environments, \toolname\ enables controlled evaluation of proof automation over repository-scale software-verification developments. To make the role of repository context explicit and measurable, we evaluate provers under two complementary regimes. The \emph{curated} regime is an intentionally oracle-like diagnostic condition, constructed using ground-truth proof dependencies to isolate context utilization from premise selection. The \emph{full-repository} regime is the more realistic deployment condition, in which the prover must identify useful declarations within the codebase.

Our experiments yield three main findings. First, provers that perform well on Mathlib-centric mathematics tasks transfer poorly to this repository-grounded setting. Second, performance degrades as \emph{transitive} dependency structure increases, such as when discharging a goal requires reasoning over a large, multi-hop closure of project-local definitions and lemmas. Third, curated context improves over full-repository context, but substantial room for improvement remains even when the relevant context is provided. This highlights that context retrieval is necessary but not sufficient for successful proof automation.

In summary, this paper makes the following contributions:
\begin{itemize}[leftmargin=*]
  \item We introduce \toolname, a benchmark of 500 Lean~4 proof obligations drawn from open-source formal-methods repositories and packaged to preserve repository context and cross-file dependencies, providing a testbed for research on repository-scale premise retrieval, noisy long-context proof generation, and reasoning over deep chains of project-specific definitions and lemmas.
  \item We provide an evaluation of frontier LLMs and specialized provers under curated and full-repository context regimes, analyzing the relationship between prover success and structural properties of repository dependence.
\end{itemize}

\section{\toolname}
\begin{table}[t]
\centering
\small
\setlength{\tabcolsep}{4pt}
\renewcommand{\arraystretch}{1.1}
\begin{tabular}{l l c c c c c}
\specialrule{1.2pt}{2pt}{2pt}
\rowcolor{gray!15}
\textbf{Proof domain} & \textbf{Benchmark} & \textbf{Lang.} &
\textbf{\makecell[l]{Proj.\\defs}} &
\textbf{\makecell[l]{Multi-\\module}} &
\textbf{\makecell[l]{Repo\\retr.}} &
\textbf{\makecell[l]{GT\\proofs}} \\
\midrule

\multirow{4}{*}{Mathematics}
 & MiniF2F & Lean & \xmark\ & \xmark & \xmark & \xmark \\
 & ProofNet & Lean & \xmark & \xmark & \xmark & \xmark \\
 & LeanDojo & Lean & \cmark\ & \cmark & \cmark & \cmark \\
 & PutnamBench & Multi. & \xmark & \xmark & \xmark & \xmark \\
\midrule

\multirow{3}{*}{\makecell[l]{Veri-\\coding}}
 & FVAPPS & Lean & \xmark & \xmark & \xmark & \xmark \\
 & CLEVER & Lean & \xmark & \xmark & \xmark & \xmark \\
 & Verina & Lean & \cmark & \xmark & \xmark & \pmark \\
\midrule

Verification
 & MiniCodeProps & Lean & \cmark & \xmark & \xmark & \xmark \\
\midrule
\midrule

\textbf{\makecell[l]{PL \& Verif.}}
 & \textbf{\makecell[l]{\toolname\\ (Ours)}} & \textbf{Lean} & \cmark & \cmark & \cmark & \cmark \\
\specialrule{1.2pt}{2pt}{2pt}

\end{tabular}
\caption{Comparison of theorem proving benchmarks. LeanDojo's scope is confined to Mathlib; \toolname\ features 23 diverse repositories.}
\label{tab:curr-bench}
\vspace{-1.5em}
\end{table}

\toolname\ is a benchmark of 500 Lean~4 proof obligations drawn from real, open-source formal-methods developments. We target Lean~4 as it is the primary language of recent neural theorem provers. As Table~\ref{tab:curr-bench} shows, existing Lean benchmarks either operate within the shared Mathlib ecosystem or introduce limited project-specific context without multi-module structure. \toolname\ fills this gap: each task is extracted from a fixed Git commit and packaged to preserve the original proof environment, including project-specific abstractions and cross-file dependencies. Theorems are phrased in terms of project-defined datatypes and invariants, and completing the proof requires reasoning within the surrounding codebase.

\subsection{Dataset Construction}

\noindent {\bf \emph{Benchmark Collection.}} The benchmark is drawn from a diverse set of 23 Lean repositories, covering compiler correctness, programming language semantics and type systems, verification frameworks, and applied verification projects (e.g., zero-knowledge circuits, protocols, and optimization problems). Figure~\ref{fig:source_category} shows the distribution of these repositories across topics. These categories reflect the current landscape of Lean~4 verification developments: unlike the Lean mathematics ecosystem, which is largely consolidated within Mathlib, software verification efforts are distributed across independent, domain-specific projects. Table~\ref{tab:detailed-repo-stats} in the appendix provides per-repository descriptions, task counts, and dependency statistics.

\noindent{\bf \emph{Benchmark Selection.}}
When selecting benchmarks, we require each task to have a \emph{valid} proof. This means that the ground-truth proof script should contain no proof holes (e.g., \texttt{sorry} or \texttt{admit}), and any lemmas or theorems it invokes are themselves recursively proven by hole-free scripts. In addition, the Lean file containing the target theorem must compile after replacing the ground-truth proof with \texttt{by sorry}, ensuring that the statement, imports, and required definitions are well-formed independently of the original proof term. Finally, we exclude trivial goals by requiring that the theorem is not discharged solely by standard automation (e.g., \texttt{simp}, \texttt{omega}, \texttt{grind}) and that it references at least one repository-local definition.

\definecolor{CComp}{HTML}{C97C7C}  
\definecolor{CAppl}{HTML}{8DA0CB}  
\definecolor{CFrame}{HTML}{7FBF9B} 
\definecolor{CSem}{HTML}{E6C46A}   
\definecolor{CType}{HTML}{C9A07A}  

\begin{wrapfigure}{r}{0.4\linewidth}
    \vspace{-1.5em}
    \centering
    \footnotesize
    \begin{tikzpicture}
    \tikzset{pieslice/.style={draw=white, line width=0.8pt}}
    \pie[
        /pgfpie/radius=1.2,
        text=legend,
        sum=auto,
        after number={},
        rotate=90,
        style={pieslice},
        color={CComp, CAppl, CFrame, CSem, CType}
    ]{
        3/Compiler,
        7/Applied verif.,
        5/Framework,
        6/Semantics,
        2/Type systems
    }
    \end{tikzpicture}
    \vspace{-0.5em}
    \caption{Repository topics.}
    \label{fig:source_category}
    \vspace{-1.5em}
\end{wrapfigure}

Beyond enforcing validity, we select tasks that span a range of difficulty metrics. To this end, we score each benchmark along two axes: \emph{proof complexity}, approximated by the length of the repository’s ground-truth tactic script, and \emph{contextual dependency}, measured by how many distinct definitions and lemma statements the proof draws from repository-local files. In summary, \toolname\ is constructed by first filtering candidate theorems for validity and non-triviality, and then sampling 500 tasks from the remaining pool to balance repository category and the two difficulty axes described above. Across repositories, we select an average of 10.4\% and a median of 8.6\% of eligible proof obligations; selection rates range from 0.2\% for lean-mlir to 26\% for LeroyCompilerVerificationCourse. Table~\ref{tab:repo-stats} summarizes the resulting dataset. Two repositories account for 119 of the 500 tasks; these are substantially larger verification developments whose full-repository context exceeds 200k tokens. We retain them deliberately, as they represent the regime where repository-scale reasoning is most challenging. Thus the non-uniform distribution reflects both the scale and maturity of the current Lean software-verification ecosystem and our goal of retaining substantial repository-scale developments. We balance diversity, realism, and evaluation cost rather than sampling uniformly by repository or exhaustively retaining every theorem; Appendix~\ref{sec:within-repo} verifies that the main trends hold within repositories.

\begin{wraptable}{r}{0.55\linewidth}
\centering
\footnotesize
\renewcommand{\arraystretch}{1.05}
\setlength{\tabcolsep}{2.5pt}
\begin{tabular}{l c c r r c r}
\specialrule{1.2pt}{2pt}{2pt}
\textbf{Cat.} & \textbf{Repos} & \textbf{Tasks} & \textbf{Curated} & \textbf{Full Repo} & \textbf{Tr.} & \textbf{Total} \\
\midrule
Regular & 21 & 381 & 5,407 & 44,934 & 21 & 813 \\
Large & 2 & 119 & 6,339 & 2,928,342 & 30 & 29,112 \\
\specialrule{1.2pt}{2pt}{2pt}
\end{tabular}
\caption{Dataset statistics. ``Large’’ = context $>$200k tokens. Values are per-repo averages.}
\label{tab:repo-stats}
\end{wraptable}

\noindent {\bf \emph{System Specific Subset.}}
Evaluation on Harmonic's Aristotle system requires strict adherence to specific library versions (Mathlib v4.24.0) and a pre-compiled project environment rather than in-prompt context presentation. To ensure compatibility, we filtered our full task set for version alignment and performed targeted manual refactoring to resolve versioning discrepancies. As the complexity of this engineering effort precluded scaling to the entire 500-task benchmark set, we collected a finalized subset of 100 fully-compatible tasks (\toolname-Aristotle), enabling evaluation on the state-of-the-art specialized prover.

Two aspects of this configuration merit clarification. First, because Aristotle operates within a compiled project, the evaluation context \emph{additionally} includes same-file lemma statements preceding the target theorem. Removing these lemmas while retaining a compilable environment would require restructuring the source files and selectively re-compiling, which was infeasible at scale. As a result, \toolname-Aristotle represents a simpler setting in which all same-file lemma statements---including those used by the ground-truth proof---are available to the prover. Second, one might ask whether the curated context (Section~\ref{subsec: context-config}) could be packaged as a standalone compilable project for Aristotle. This is not straightforward: the curated context tracks definitions transitively but does not transitively track the lemma dependencies of included lemmas, so it does not in general form a self-contained, compilable unit.

\begin{figure*}[t]
    \centering
    \includegraphics[width=0.94\textwidth, trim=300 300 300 50, clip]{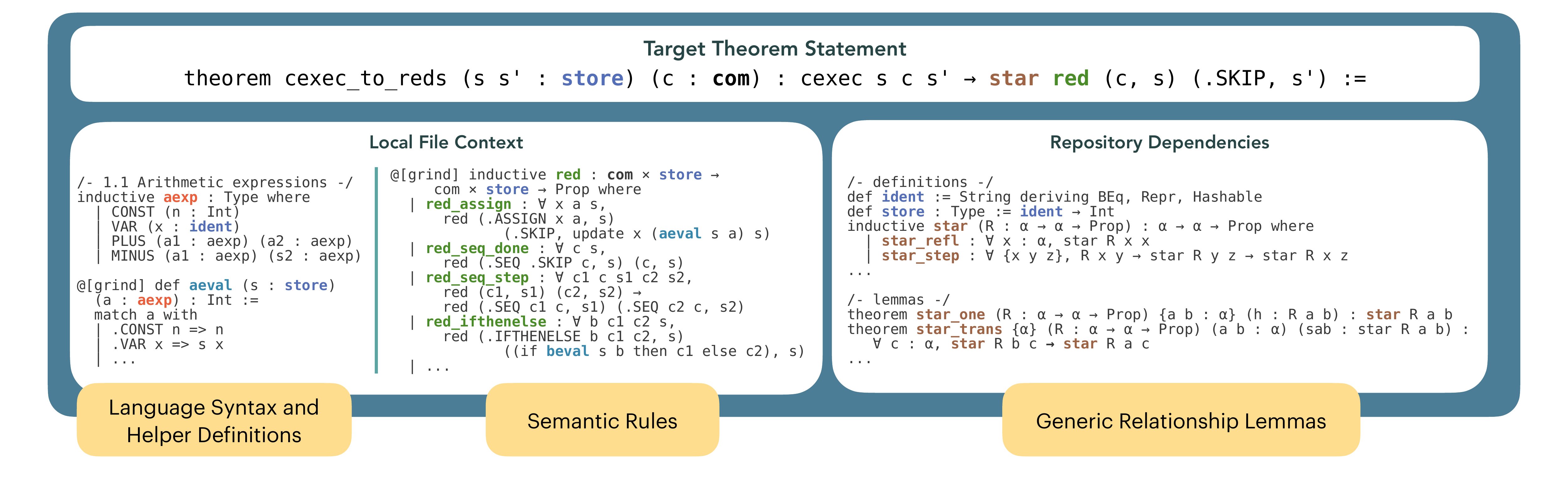}
    \caption{An example \toolname\ task. The target \texttt{cexec\_to\_reds} relates two project-defined execution relations. The left (local-file) panel supplies the language syntax and semantic rules in scope at the theorem; the right (repository) panel supplies cross-file base types and relational lemmas. Matching colors trace each project-specific symbol in the target to the declaration that gives it meaning, illustrating why the proof cannot be treated as an isolated theorem statement.}
    \label{fig:task}
    \vspace{-1.5em}
\end{figure*}

\noindent {\bf \emph{Task Composition.}}
Each task consists of the target theorem, the local file context up to the theorem, all required local-repository definitions and helpful repository lemma declarations for stating and proving the target theorem
, and the necessary library imports, as illustrated in Figure~\ref{fig:task}.
Tasks are extracted without rewriting definitions or replacing local abstractions with equivalent standard-library formulations, preserving the original dependency structure of each proof.

\subsection{Contextual Configurations}
\label{subsec: context-config}

To study how provers use repository-local context under different levels of noise, we evaluate each task under two contextual configurations. Both configurations include the same \emph{base context}, which includes: (i) the contents of the local file up to the target theorem (i.e., the definitions in scope) excluding the lemmas, and (ii) the external library (e.g., Mathlib, Std) imports and declarations required for the theorem statement to type-check.

\noindent {\bf \emph{Curated Context.}}  
The curated context mode is an intentionally oracle-like diagnostic condition that provides the model with a focused set of dependencies designed to approximate the necessary background knowledge with minimal distraction for completing a proof.
In addition to the base context, it includes only the explicit repository dependencies: the specific definitions and lemmas from other files in the same repository that are directly invoked by the ground truth target theorem, and the transitively invoked definitions by any dependency in the context. To ensure the integrity of the evaluation and prevent models from mimicking domain-specific proof scripts, we elide the proof bodies of all provided lemmas and definitions. Because these premises are identified from the ground-truth proof, this configuration is not representative of deployment. It instead represents a best-case diagnostic scenario that separates context utilization from retrieval performance without leaking the target proof.


\noindent {\bf \emph{Full  Context.}} In this setting, the prover receives the entire repository library in addition to the base context. This includes all definitions and lemma/theorem statements across \emph{all files} in the  repository.   It is worth noting that the resulting context can be extremely long: among the 500 tasks, 119 come from repositories whose full-repository context exceeds hundreds of thousands of tokens, and one reaches hundreds of millions of tokens (Table~\ref{tab:repo-stats}). Such instances are far beyond the effective context windows of current models, underscoring the need for reliable context-selection mechanisms. When the full repository context exceeds a model's context window, we first ensure that all declarations present in the curated context are included, then fill the remaining window capacity with additional repository declarations. This guarantees that the relevant dependencies are always available, even when the full repository cannot be provided in its entirety.

\paragraph  {\bf \emph{Proof isolation.}}
Both configurations are designed to ensure that the target proof is never revealed in the provided context: we include only \emph{preceding} content within the local file and never provide subsequent lemmas or proofs. Whenever we provide repository lemmas or definitions (either in the curated set or as part of the full repository), we include only their statements and elide proof bodies. Nevertheless, the Full Context configuration exposes the broader module ecosystem in which the theorem lives. As such, this ambient structural information can provide cues about recurring abstraction patterns or common proof structure that is largely absent in the curated configuration.

\section{Experiments}
\subsection{Baselines}

To evaluate performance on \toolname, we select a diverse set of models spanning frontier general-purpose LLMs and specialized formal reasoning systems. 

We evaluate three primary baseline models that represent the current frontier of general-purpose reasoning and code generation: GPT-5.2, Claude-Opus-4.5, and Gemini-3-Pro. To broaden the evaluation beyond closed frontier models, we include Goedel-Prover-V2-32B~\citep{goedelv2}, an open-weight specialized Lean prover. We also evaluate Harmonic's Aristotle Prover~\citep{aristotle}, a strong specialized prover system. These systems cover complementary categories---general-purpose LLMs, an open-weight prover model, and an integrated prover service---while remaining compatible with independently packaged Lean repositories and their project-specific toolchains.


\subsection{Evaluation pipeline}
\begin{wrapfigure}{r}{0.5\linewidth}
  \centering
  \includegraphics[width=\linewidth, trim=50 80 520 300, clip]{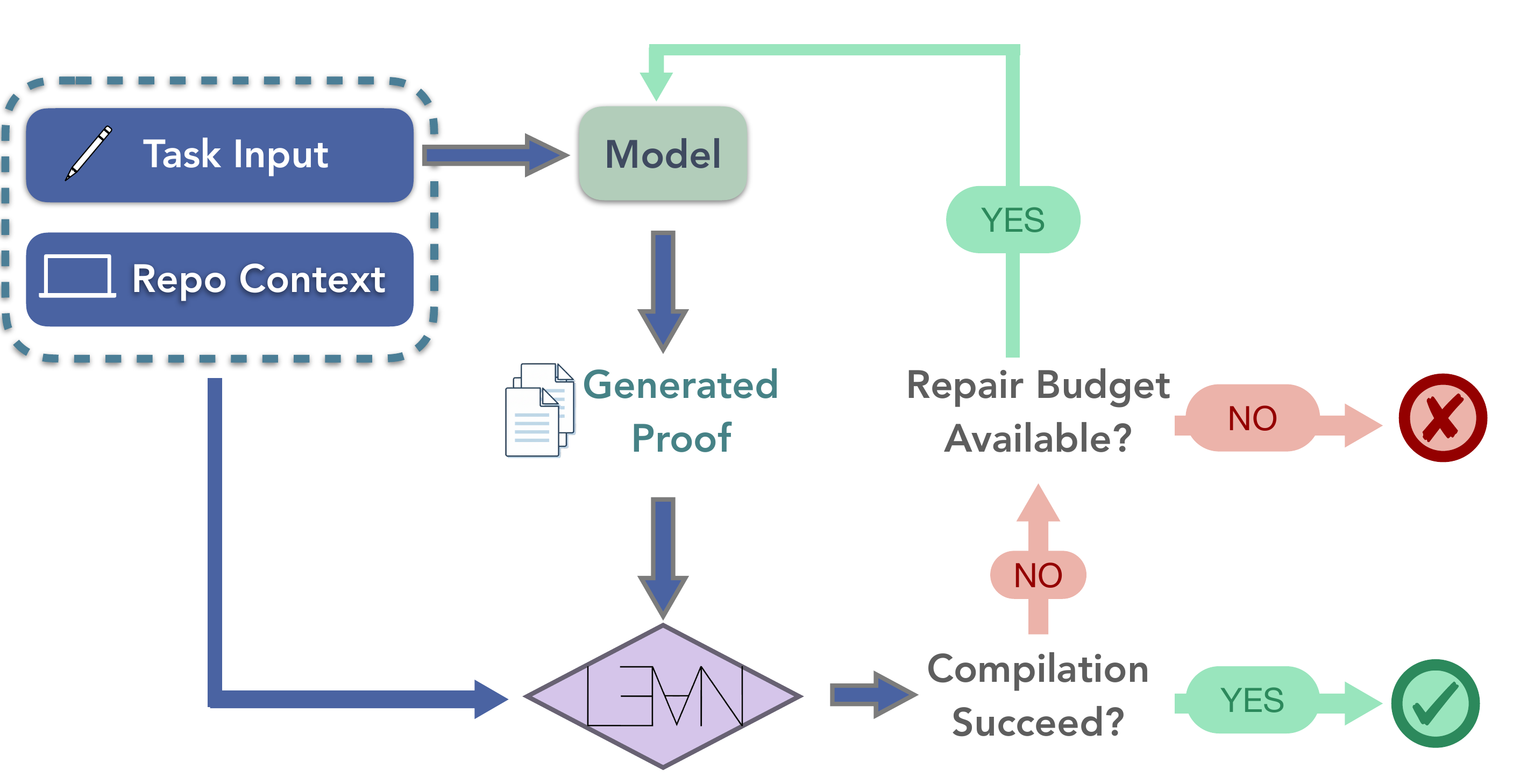}
  \caption{Evaluation pipeline.}
  \label{fig:pipeline}
  \vspace{-3em}
\end{wrapfigure}

We evaluate general-purpose LLMs using a  generation-check-repair loop, as illustrated in Figure~\ref{fig:pipeline}. For each evaluation run, the model receives a task associated with the selected contextual configuration as input, and the evaluation proceeds in three stages: (1) we sample $k$ candidate proof scripts from the model. Each candidate is generated as a tactical proof block intended to satisfy the target goal; (2) we wrap each generated script into its original file context and attempt verification using the Lean 4 compiler. To ensure a stable and efficient evaluation, the all task environments are pre-built prior to verification; we use the specific toolchain version associated with the source repository of each task and enforce a 300-second compilation timeout per candidate\footnote{Large dependencies such as Mathlib are compiled once and cached per repository toolchain version.}; (3) for any failing script, we invoke the model for up to $r$ repair rounds using the compiler error message, corresponding code lines from each error location, and the current proof state as feedback. A task is marked as ``Solved'' if any of the $k$ candidates or their subsequent repairs pass verification. For {specialized theorem-proving systems} with an internal feedback loop (e.g., Goedel-Prover-V2 and Aristotle), we evaluate them as end-to-end systems rather than wrapping them in an external repair loop, as they already contain their own internal feedback loop.

\subsection{Data Configurations}\label{sec:data-config}

We consider two variations of \toolname\  that differ in scope and in the degree to which they accommodate prover-specific engineering constraints.

\paragraph{\toolname-Full.}
Our primary evaluation setting is the full \toolname\ benchmark, consisting of 500 proof obligations. This setting is used for all models and provers that can consume our standard packaging of repository context.

\paragraph{\toolname-Aristotle.}
As described in Section~3.1, Aristotle requires a pre-compiled project environment rather than in-prompt context. We evaluate it on the 100-task \toolname-Aristotle subset, which additionally includes same-file lemma statements (see Section~3.1 for details on why this is necessary and why the curated context cannot serve as a substitute). To enable a controlled comparison on this subset, we also evaluate Gemini-3-Pro under the same configuration.


\section{Results and Analysis}
\begin{table}[t]
\centering
\small
\setlength{\tabcolsep}{4pt}
\renewcommand{\arraystretch}{1.1}
\begin{tabular}{l l c c}
\specialrule{1.2pt}{2pt}{2pt}
 & & \multicolumn{2}{c}{\textbf{Context Conditions}}\\
\cmidrule(lr){3-4}
\multirow{-2}{*}{\textbf{Data Setting}} & \multirow{-2}{*}{\textbf{Model Setting}} & \textbf{Curated} & \textbf{Full Repo} \\
\midrule

\multirow{4}{*}{\toolname-Full}
 & Claude Opus 4.5 (Pass@8, $r\!=\!3$) & $31.2\%$ &  $23.2\%$ \\
 & GPT-5.2 (Pass@8, $r\!=\!3$) & $12.6\%$ & $10.8\%$ \\
 & Gemini-3-Pro (Pass@8, $r\!=\!3$)  & $41.0\%$  & $34.8\%$  \\
& Goedel-Prover-v2 (Pass@8, $r\!=\!0$) & $5.6\%^\dagger$ & $5.2\%^\dagger$  \\
\midrule

\multirow{2}{*}{\makecell[l]{\toolname-Aristotle}}
 & Aristotle  & - & $69\%^*$   \\
 & Gemini-3-Pro (Pass@8, $r\!=\!3$) & - & $65\%^*$ \\
\specialrule{1.2pt}{2pt}{2pt}
\end{tabular}
\caption{\toolname\ evaluation results. $^\dagger$ Four tasks whose required context exceeds the Goedel-Prover-V2 context window are excluded from both context conditions. $^*$Uses the separate 100-task compatibility subset with full context and preceding same-file lemma statements (Section~\ref{subsec: context-config}); these rows are not directly comparable to \toolname-Full.}
\label{tab:detailed-results}
\vspace{-1em}
\end{table}

\noindent {\bf \emph{Model Performance.}}
Table~\ref{tab:detailed-results} summarizes the verification success rates across our two data settings (Full Set, Subset) and two contextual configurations (Curated Context, Full Context). Overall, performance is low, highlighting the inherent difficulty of the \toolname\ tasks. 

\begin{wrapfigure}{r}{0.44\linewidth}
    \vspace{-0.8em}
    \centering
    \includegraphics[width=\linewidth]{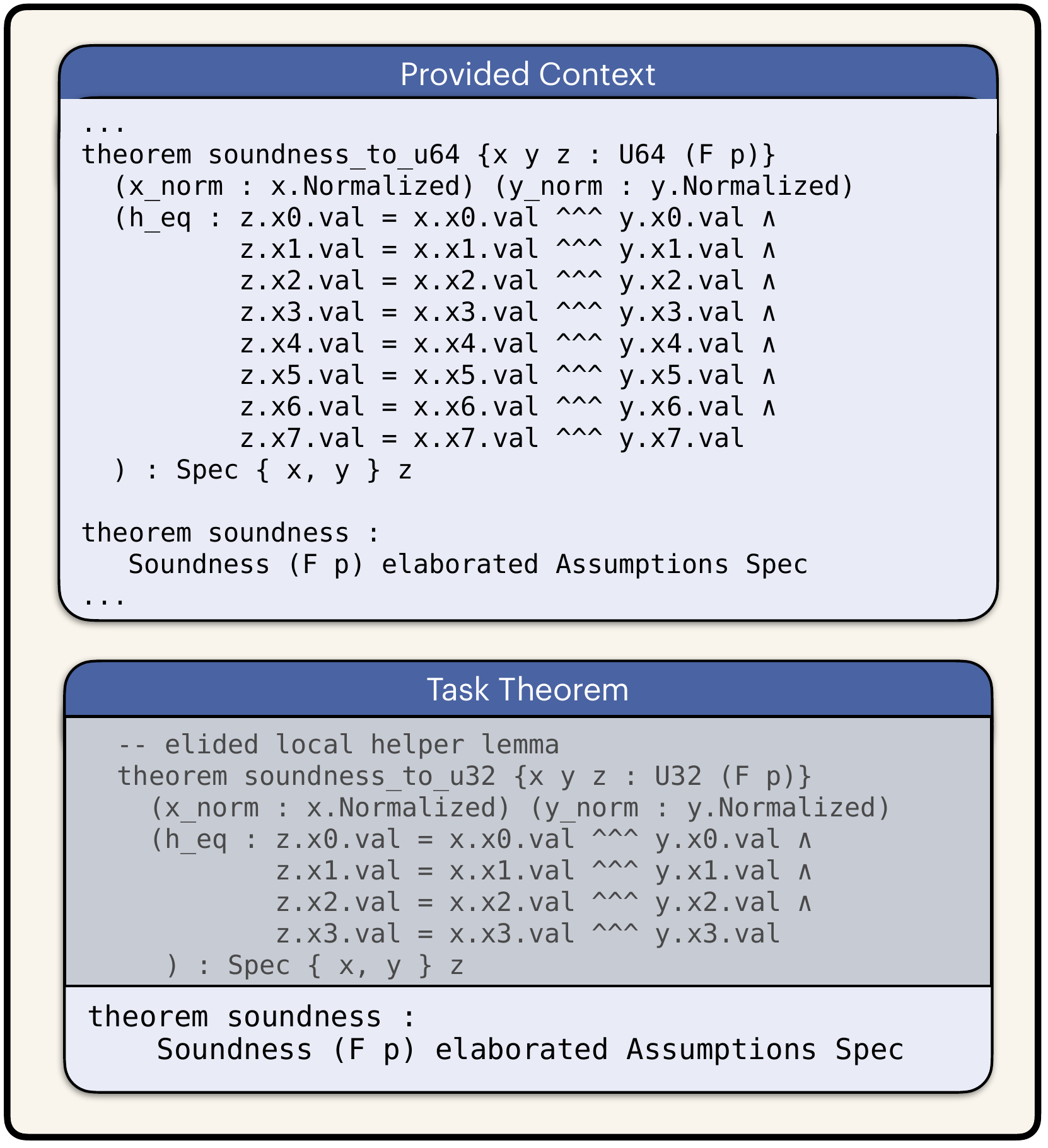}
    \vspace{-0.6em}
    \caption{Full-repository context exposes the analogous \texttt{soundness\_to\_u64} proof as a pattern for the target \texttt{soundness\_to\_u32}; curated context omits it.}
    \label{fig:full-context}
    \vspace{-0.8em}
\end{wrapfigure}

Under the Curated Context, the best-performing model, Gemini-3-Pro, achieves a success rate of $41.0\%$, followed by Claude Opus 4.5 at $31.2\%$. Thus, even oracle-like dependency selection leaves 59\% of tasks unsolved, showing that retrieval is not the only bottleneck. In the more challenging Full Repo Context, success rates decrease across all evaluated models; Gemini-3-Pro and Claude Opus 4.5 achieve $34.8\%$ and $23.2\%$, respectively, while GPT-5.2 reaches $10.8\%$. The open-weight Goedel-Prover-V2 solves $5.6\%$ under curated context and $5.2\%$ under full-repository context, indicating limited transfer to repository-scale software verification.


\noindent {\bf \emph{Aristotle Comparison.}}
Aristotle requires a compiled project environment, so we evaluate it separately on the 100-task \toolname-Aristotle subset and compare it with Gemini-3-Pro under the same configuration (Section~\ref{subsec: context-config}). With preceding same-file lemma statements available, Aristotle solves $69\%$ of the tasks and Gemini-3-Pro solves $65\%$. Removing these statements and evaluating Gemini-3-Pro under the standard full-context protocol reduces its success rate to $41\%$, showing that the shared configuration is substantially easier than the main benchmark setting. The four-point difference in the matched comparison captures the combined effect of Aristotle's compiled environment and agentic proof search.

Performance also varies markedly by repository scale. On the 21 regular-sized repositories (381 tasks), Gemini-3-Pro achieves $47.6\%$ (curated) and $43.7\%$ (full). On the two large repositories whose full context exceeds 200k tokens (ArkLib and lean-mlir, 119 tasks), success drops to $17.6\%$ and $10.9\%$. For these large repositories, the full context far exceeds model context windows; as described in Section~\ref{subsec: context-config}, we prioritize curated dependencies and fill the remaining window with additional repository declarations.



\noindent {\bf \emph{Comparison of Context Conditions.}}
Full-repository context consistently reduces success, although by less than its much larger search space might suggest (Table~\ref{tab:detailed-results}). For Gemini-3-Pro, for example, the gap between curated and full-repository context is 6.2 percentage points (41.0\% vs.\ 34.8\%). Curated context uses ground-truth proof dependencies to identify relevant declarations, whereas full-repository context requires the prover to find them among many additional declarations. The gap therefore measures the effect of this added repository context while keeping the relevant dependencies available in both conditions.

One possible explanation is that full context provides compensating signals despite its much larger search space. Files outside the ground-truth dependency closure can contain structurally similar proofs, auxiliary lemmas, or recurring abstraction patterns. Figure~\ref{fig:full-context} gives a concrete example: \texttt{soundness\_to\_u64} is not required by the target proof and is therefore absent from curated context, but its proof can cue the construction of \texttt{soundness\_to\_u32}. This is an illustrative mechanism rather than a causal attribution for the aggregate gap.

\begin{figure}[t]
    \centering
    \begin{minipage}[t]{0.54\linewidth}
        \vspace{0pt}
        \centering
        \includegraphics[width=\linewidth]{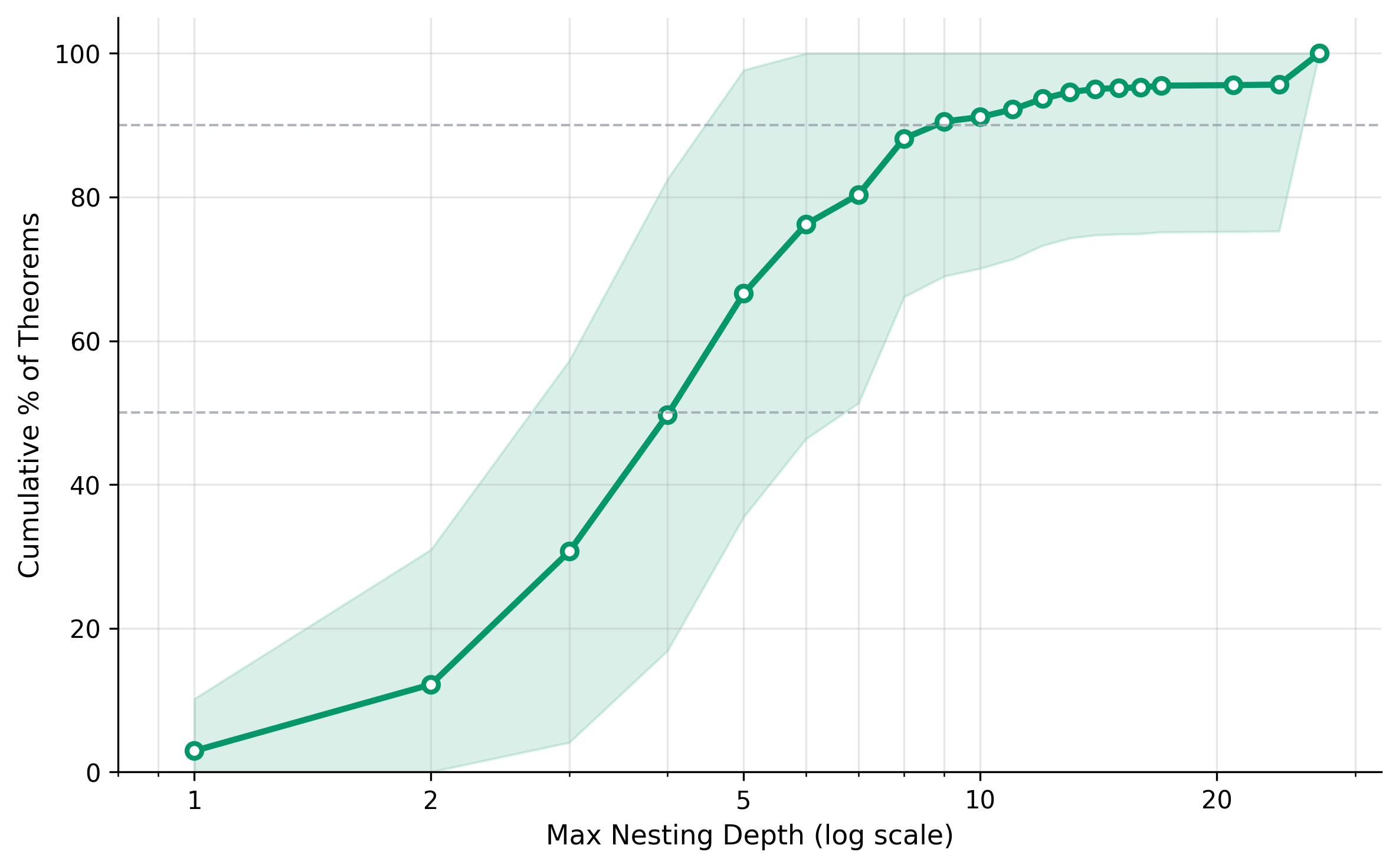}
        \vspace{-0.8em}
        \captionof{figure}{Cumulative distribution of dependency depth (log scale). Line: mean; shading: $\pm$1 std.}
        \label{fig:nested_dependency}
    \end{minipage}\hfill
    \begin{minipage}[t]{0.43\linewidth}
        \vspace{0pt}
        \centering
        \footnotesize
        \setlength{\tabcolsep}{2pt}
        \renewcommand{\arraystretch}{1.0}
        \resizebox{\linewidth}{!}{%
        \begin{tabular}{l c c c c c}
            \specialrule{1.2pt}{2pt}{2pt}
            \rowcolor{gray!20}
            & & \makecell{Lib \\ (Dir.)} & \makecell{Lib \\ (Tr.)} & \makecell{Proj \\ (Dir.)} & \makecell{Proj \\ (Tr.)} \\
            \rowcolor{gray!5}
            \textbf{Bench.} & \textbf{N} & & & & \\
            \midrule
            Ours   & 500 & 2.3 & 11.2 & 11.6 & 37.9 \\
            Verina & 189 & 0.0 & 0.0 & 3.0 & 3.0 \\
            Putnam & 662 & 3.3 & 3.3 & 0.0 & 0.0 \\
            \specialrule{1.2pt}{2pt}{2pt}
        \end{tabular}}
        \captionof{table}{Dependency footprint. Lib = library; Proj = project (excluding same-file); Dir.\ = direct; Tr.\ = transitive.}
        \label{tab:depedency_comparison}
    \end{minipage}
    \vspace{-1em}
\end{figure}

\noindent {\bf \emph{Impact of High Contextual Dependency.}} The difficulty of our benchmark is closely tied to its unusually high degree of contextual dependency. Table \ref{tab:depedency_comparison} quantifies the size of the dependency footprint. Theorems in \toolname\ directly reference only a modest number of items (2.3 library and 11.6 project dependencies on average), but their \emph{transitive closure} expands to 11.2 library dependencies and 37.9 project dependencies \emph{per theorem}. In contrast, Verina tasks involve only shallow local dependencies (around 3 transitive dependencies on average), while PutnamBench theorems are entirely self-contained with zero project dependencies.

However, dependency size alone does not fully capture structural difficulty. Figure~\ref{fig:nested_dependency} shows the dependency \emph{nesting depth} --- i.e., the length of the longest chain in the transitive dependency graph induced by referenced definitions and lemmas.  The distribution reveals that many theorems require reasoning across long dependency chains: roughly half of the theorems have nesting depth $\geq 5$, and around 10\% exceed depth 12. This indicates that relevant information is often separated from the theorem statement by multiple layers of intermediate abstractions.

These two observations highlight an important distinction. A theorem may depend on dozens of definitions (large dependency size), but the key difficulty arises when those dependencies are arranged in long chains, where each step builds on the previous one. Our correlation analysis (Table~\ref{tab:correlation}, Appendix) confirms this: nesting depth shows the strongest correlation with failure (Spearman $r = -0.356$, full), followed by total transitive dependency count ($r = -0.337$), while direct dependency counts are more weakly correlated. We report correlations for additional features including proof length in the Appendix.

To make this trend concrete, we bucket tasks by dependency depth and report success rates for Gemini-3-Pro (with repair) under each context regime. At depth 1--3, success is $38.0\%$ (curated) and $32.6\%$ (full); at depth 4--6, $25.5\%$ and $23.5\%$; at depth 7+, $21.1\%$ and $15.6\%$. The decline is monotonic across both regimes, and the gap between curated and full context widens at higher depths. We note that our benchmark is intentionally weighted toward tasks with nontrivial dependency structure (50\% at depth $\geq 5$), as this regime is underrepresented in existing benchmarks and reflects the structure of real verification codebases. Tasks with shallow dependencies are already well served by benchmarks such as Verina and MiniCodeProps.


\noindent {\bf \emph{Comparison of Proof Patterns.}}  Beyond dependency structure, \toolname\ also differs from prior datasets in the proof patterns it induces. To compare proof styles across benchmarks in a controlled way, ground-truth proofs are not sufficient: many benchmarks do not provide them, and even when available, human-written proofs and prover-specific proof search can introduce stylistic confounders. We therefore perform a cross-dataset comparison (e.g., Verina vs.\ \toolname) using Gemini-3-Pro generated proofs that successfully verify in both datasets. This holds the prover constant and allows observed differences to be attributed primarily to the theorems themselves.

Under this controlled comparison, Verina and \toolname\ exhibit systematically different tactic profiles. In Verina, 45\% of correct Gemini-generated proofs open with \texttt{unfold} and 34\% close with \texttt{simp}, with automation tactics accounting for 29\% of tactic usage overall; 13\% of Verina proofs are discharged by a single automation call. \toolname\ proofs follow a markedly different pattern: only 7\% begin with \texttt{unfold}, 10\% close with \texttt{simp}, and none are solved by one-shot automation. Automation accounts for just 10\% of tactic usage in \toolname\ proofs, while 71\% exhibit branching structure involving case analysis or induction (compared to 38\% in Verina). These differences suggest that Verina proofs can often be resolved by unfolding definitions and invoking automation, whereas \toolname\ tasks require multi-step reasoning with explicit case splits and lemma application.

\section{Related Work}

\noindent {\bf \emph{Formal Verification and Benchmarks.}}
Formal verification provides strong assurance but proof effort remains a bottleneck. Tools such as Dafny~\citep{dafny} and Verus~\citep{verus} automate proofs via SMT solvers, while interactive theorem provers support manual proof construction, enabling verified artifacts such as CompCert~\citep{compCert}, seL4~\citep{sel4}, and CakeML~\citep{cakeML}. Domain-specific frameworks like Iris~\citep{iris}, Verdi~\citep{verdi}, and Loom~\citep{loom} reduce overhead but still require significant manual effort.
On the benchmarking side, early suites such as HumanEval~\citep{human-eval} and MBPP~\citep{mbpp} validate code by unit tests without formal guarantees. DafnyBench~\citep{dafnyBench} targets invariant synthesis and MiniCodeProps~\citep{minicodeprops} evaluates proofs for small functional programs, while FVAPPS~\citep{fvapps}, CLEVER~\citep{clever}, and Verina~\citep{verina} target verified code generation. \toolname\ complements these by focusing on proof obligations embedded in repository-scale Lean developments with substantial project-specific context.

\noindent {\bf \emph{General ITP Benchmarks.}}
Widely used Lean benchmarks target formalized mathematics: MiniF2F~\citep{minif2f}, ProofNet~\citep{proofNet}, LeanDojo~\citep{leanDojo}, and PutnamBench~\citep{putnamBench} operate within a shared Mathlib ecosystem. The miniCTX effort~\citep{minictx} supplies surrounding project context but remains focused on mathematical theorems. Outside Lean, CoqGym~\citep{astactic} aggregates 71K proofs from 123 Coq projects but targets step-by-step tactic prediction, where the proof assistant manages the environment at each goal, rather than the whole-proof generation paradigm used by current LLM-based provers. \toolname\ complements CoqGym by targeting whole-proof generation in Lean~4 with explicitly controlled repository context. Extending similar benchmarks to other ITPs (e.g., Rocq, Isabelle) is a natural direction for future work.

\noindent {\bf \emph{LLM-based Prover Systems.}} Neural proof automation has progressed from early tactic prediction~\citep{htps,fm-scl,gpt-f} to agentic systems using verifier feedback. COPRA~\citep{copra} and AxProver~\citep{ax-prover} run generate-check-repair loops; DSP~\citep{dsp} constructs intermediate sketches. For premise selection, LeanDojo~\citep{leanDojo} and Rango~\citep{rango} apply retrieval-augmented proving. Specialized prover models include Llemma~\citep{llemma}, Kimina-Prover~\citep{kimina}, DeepSeek-Prover-V2~\citep{deepseek-prover-v2}, G\"odel-Prover-V2~\citep{goedelv2}, SeedProver~\citep{seedprover}, and Hilbert-Prover~\citep{hilbert}. Despite significant advances, these systems are predominantly evaluated on mathematics-centric benchmarks, motivating \toolname.

\section{Limitations and Future Work}
Our benchmark evaluates single-theorem proving for fully specified goals, and therefore does not capture multi-stage verification workflows such as specification synthesis, proof planning, or joint proving of multiple theorems where intermediate lemmas serve several goals. Even within this setting, we focus on native long-context reasoning rather than evaluating practical retrieval-augmented or agentic context-selection baselines. The curated regime is an oracle-like diagnostic condition, not a deployable retrieval method: it uses ground-truth proof dependencies to separate premise selection from context utilization. Its best result of 41.0\% shows substantial headroom even after relevant premises have been identified, but does not predict the performance of a practical retriever. \toolname\ is intended to support follow-up work on repository-scale retrieval and premise selection, long-context proof generation under noisy context, and robustness to multi-hop project-specific dependencies. Our evaluation also does not isolate why Goedel-Prover-V2 transfers poorly to this setting: project-specific abstractions, long dependency chains, distribution shift from mathematics benchmarks, and the Pass@8 budget may all contribute. Finally, because \toolname\ draws from public GitHub repositories, we cannot fully rule out contamination in the pre-training data of closed-weight models. That said, the strong correlation between dependency depth and failure suggests that memorization alone is insufficient for these tasks.

\section{Conclusion}

We introduced \toolname, a benchmark contribution comprising 500 Lean~4 proof obligations from open-source formal-methods developments while preserving project context and cross-file dependencies. Our experiments show that current provers achieve only modest success, with performance degrading most strongly as dependency nesting depth and transitive dependency structure grow. The intentionally oracle-like curated condition improves performance but still leaves substantial headroom, indicating that repository-scale verification requires both effective context selection and reasoning over project-specific abstractions. We hope \toolname\ supports future experiments on retrieval and premise selection, noisy long-context generation, and robust reasoning across deep repository dependency chains.


\section*{Acknowledgments}
We thank the anonymous reviewers for their helpful feedback. This work was conducted in a research group supported by NSF awards CCF-1918889, CNS-2120696, CCF-2210831, and CCF-2319471, CCF-2422130, CCF-2403211, CCF-2505865, CCF-2326576, and CCF-2403211 as well as a DARPA award under agreement HR00112590133 and HR0011262E020.
This work is also supported in part by a Canada CIFAR AI Chair award to QC. 
This research computation has been supported by the
Vista GPU Cluster through the Center for Generative AI (CGAI) and the Texas Advanced
Computing Center (TACC) at the University of Texas at Austin, as well as a Google Compute Grant for running Gemini-3-Pro. 


\bibliography{colm2026_conference}
\bibliographystyle{colm2026_conference}

\clearpage
\appendix
\section{Prompts}

\lstset{
    basicstyle=\ttfamily\small,
    breaklines=true,            
    breakatwhitespace=false,     
    literate={-}{{-}}1          
}

\newtcolorbox{promptbox}[1]{
    enhanced,
    breakable,                 
    colback=green!10!white,    
    colframe=black,            
    colbacktitle=black,        
    title={#1},                
    fontupper=\ttfamily, 
    arc=2mm,                   
    boxrule=0.5pt,
    left=5pt, right=5pt, top=5pt, bottom=5pt
}

\begin{promptbox}{Prompt for Proof Generation (Curated Context Mode)}
\begin{lstlisting}
You are an expert Lean4 and formal verification agent proving a Lean4 theorem. Your output **must** strictly adhere to the specified format.

### Input Information
You will receive the following components, relevant to the target theorem:
* <used_lib_defs>...</used_lib_defs>: Imported library definitions directly used by the theorem.
* <used_repo_defs>...</used_repo_defs>: Definitions from the local project (but outside the current file), directly or transitively used by the theorem.
* <lib_lemmas>...</lib_lemmas>: Imported library lemmas/theorems that may assist the proof.
* <repo_lemmas>...</repo_lemmas>: Lemmas/theorems defined in the local project outside this file.
* <local_ctx>...</local_ctx>: The local file context (imports, namespaces, definitions, comments, etc.) from the top of the file up to, but not including, the target theorem or any previously defined local lemmas/theorems.
* <target_theorem>...</target_theorem>: The target theorem statement, ending with `:=`.

**Notes**
1. Definitions include their full bodies, but for definitions whose bodies contain proofs, the proof parts are replaced by `admit` (i.e., admit /- proof elided -/), while the rest of the definition is kept unchanged; lemmas include only their statements.
2. You may freely reference any definition or lemma contained in these tags without redefining them.
3. The tags for used definitions and helpful lemmas (used_lib_defs, used_repo_defs, lib_lemmas, repo_lemmas) may be omitted if empty.

### Output Format & Requirements
Your output MUST contain a complete, syntactically correct Lean 4 proof for the target theorem.

**Strict Output Structure:**
<reasoning> 
...
</reasoning> 

<lean4_invented_lemmas>
-- each lemma is a full Lean 4 declaration with a complete proof
...
</lean4_invented_lemmas>

<lean4_proof>
-- Lean 4 proof body of the target theorem
...
</lean4_proof>

**Requirements:**
1. <reasoning> and <lean4_invented_lemmas> are OPTIONAL. Include them only if they aid proof construction.
2. <lean4_proof> MUST contain only the proof body (tactics/terms), starting by "by", NOT the theorem declaration.
3. <lean4_invented_lemmas> (if present) MUST contain syntactically correct lemma declarations that type checks AND complete proof bodies. Defining a lemma using `axiom` is NOT allowed.
4. <reasoning> contains high-level proof planning.
\end{lstlisting}
\end{promptbox}

\begin{promptbox}{Prompt for Proof Generation (Curated Context Mode) - Cont}
\begin{lstlisting}
5. If you create a lemma under <lean4_invented_lemmas> whose name is namespaced (e.g., X.Y.Z), you MUST always refer to it using its fully qualified name. Do NOT rely on shortened names. For example, write rw [X.Y.Z], not rw [Z] for invented lemmas.
6. All Lean 4 code MUST be syntactically correct. No `sorry`, `admit`, or non-Lean text. Comments allowed using `--` or `/- ... -/`.
7. Follow the tags and structure EXACTLY. No extra text before, between, or after the sections.
8. Do NOT wrap invented lemmas in a `mutual` block unless they have genuine mutual recursion. Declare each lemma independently.

### Examples are provided here (example input are under "User Prompt", and example assistant response are under "Agent Response"):
{}

### Your task:
{}  
\end{lstlisting}

\end{promptbox}

\newtcolorbox{promptfull}[1]{
    enhanced,
    breakable,
    colback=pink!10!white,    
    colframe=black,            
    colbacktitle=black,        
    title={#1},                
    fontupper=\ttfamily, 
    arc=2mm,                   
    boxrule=0.5pt,
    left=5pt, right=5pt, top=5pt, bottom=5pt
}

\begin{promptfull}{Prompt for Proof Generation (Full Context Mode)}
\begin{lstlisting}
You are an expert Lean4 and formal verification agent proving a Lean4 theorem. Your output **must** strictly adhere to the specified format.

### Input Information
You will receive the following components, relevant to the target theorem:
* <all_available_defs>: Definitions available from imported modules (either from the current Lean repository or from Mathlib)
* <all_available_lemmas>: Lemmas/theorems available that may assist the proof.
* <local_ctx>...</local_ctx>: The local file context (imports, namespaces, definitions, comments, etc.) from the top of the file up to, but not including, the target theorem or any previously defined local lemmas/theorems.
* <target_theorem>...</target_theorem>: The target theorem statement, ending with `:=`.

**Notes**
1. Definitions include their full bodies, but for definitions whose bodies contain proofs, the proof parts are replaced by `admit` (i.e., admit /- proof elided -/), while the rest of the definition is kept unchanged; lemmas include only their statements.
2. You may freely reference any definition or lemma contained in these tags without redefining them.
3. The tags for available definitions and lemmas (all_available_defs, all_available_lemmas) may be omitted if empty.

### Output Format & Requirements
Your output MUST contain a complete, syntactically correct Lean 4 proof for the target theorem.

**Strict Output Structure:**
<reasoning> 
...
</reasoning> 

<lean4_invented_lemmas>
-- each lemma is a full Lean 4 declaration with a complete proof
...
</lean4_invented_lemmas>

<lean4_proof>
-- Lean 4 proof body of the target theorem
...
</lean4_proof>

**Requirements:**
1. <reasoning> and <lean4_invented_lemmas> are OPTIONAL. Include them only if they aid proof construction.
2. <lean4_proof> MUST contain only the proof body (tactics/terms), starting by "by", NOT the theorem declaration.
3. <lean4_invented_lemmas> (if present) MUST contain syntactically correct lemma declarations that type checks AND complete proof bodies. Defining a lemma using `axiom` is NOT allowed.
4. <reasoning> contains high-level proof planning.
5. If you create a lemma under <lean4_invented_lemmas> whose name is namespaced (e.g., X.Y.Z), you MUST always refer to it using its fully qualified name. Do NOT rely on shortened names. For example, write rw [X.Y.Z], not rw [Z] for invented lemmas.
6. All Lean 4 code MUST be syntactically correct. No `sorry`, `admit`, or non-Lean text. Comments allowed using `--` or `/- ... -/`.
7. Follow the tags and structure EXACTLY. No extra text before, between, or after the sections.
8. Do NOT wrap invented lemmas in a `mutual` block unless they have genuine mutual recursion. Declare each lemma independently.
\end{lstlisting} 
\end{promptfull}

\begin{promptfull}{Prompt for Proof Generation (Full Context Mode) - Cont}
\begin{lstlisting}
### Examples are provided here (example input are under "User Prompt", and example assistant response are under "Agent Response"):
{}

### Your task:
{}
\end{lstlisting}
\end{promptfull}

\newtcolorbox{promptbox2}[1]{
    enhanced,
    breakable,
    colback=blue!10!white,    
    colframe=black,            
    colbacktitle=black,        
    title={#1},                
    fontupper=\ttfamily, 
    arc=2mm,                   
    boxrule=0.5pt,
    left=5pt, right=5pt, top=5pt, bottom=5pt
}

\begin{promptbox2}{Prompt for Proof Repair}
\begin{lstlisting}
You are an expert Lean 4 and formal verification agent fixing broken Lean 4 code. Your output **must** strictly adhere to the specified format.

### Input Information
You will receive the following components:
* <code_before_error>...</code_before_error>: supporting Lean 4 context, including import statements, variables, and any definitions that precede the broken code.
* <potential_broken_lemmas>...</potential_broken_lemmas>: one or more lemma declarations and their proofs that may contain errors. This section is optional.
* <potential_broken_theorem>...</potential_broken_theorem>: the theorem declaration and proof that may contain errors. Exactly one theorem will be provided here.
* <err_msg>...</err_msg>: one or more error messages. Each error will be wrapped in <err_<id>>...</err_<id>> tags. The content inside is structured as: <Code snippet where the error occurs>\n ErrorMsg: <actual message from the Lean compiler> 

*Repair Requirements*
1. Identify where and why the error occurs based on the text given in each error message.
2. If an error is in a lemma, you may modify both the lemma statement (declaration) and its proof.
3. If an error is in the theorem, you MUST NOT change the theorem statement; you may only modify its proof.
4. The modification should be the minimal, most idiomatic Lean 4 change necessary to resolve the error.
5. You must prioritize solving the errors instead of avoiding them.

### Output Format & Requirements
Your output MUST contain Lean 4 code that is syntactically correct.

**Strict Output Structure:**
<fixed_lean4_lemmas>
-- All candidate lemmas: each contains a full statement/declaration and proof.
</fixed_lean4_lemmas>

<fixed_lean4_thm_proof>
-- Lean 4 proof body (term or tactic block) for the potentially broken theorem,
-- WITHOUT repeating its declaration.
</fixed_lean4_thm_proof>

**Output Requirements:**
1. Omit a section entirely if you make no changes to that component. E.g., if all lemmas are correct, omit <fixed_lean4_lemmas>. If the theorem proof is correct, omit <fixed_lean4_thm_proof>
2. `<fixed_lean4_lemmas>` (if present) MUST containfull, type-checking Lean 4 lemma declarations and complete proofs.
   - You MUST output every candidate lemma provided in <potential_broken_lemmas>
3. `<fixed_lean4_thm_proof>` (if present) MUST contain only the proof body (term or tactics), NOT the theorem declaration, because you are NOT allowed to modify the theorem declaration.
4. Follow the tags and structure EXACTLY. No extra text before, between, or after the sections.
5. The output MUST NOT use keywords that skip the proof (e.g., admit, sorry). 

Here is your error fixing task:
{}
\end{lstlisting}
\end{promptbox2}

\section{Repository Statistics and Dependency Analysis}

Table~\ref{tab:detailed-repo-stats} provides per-repository statistics for all 23 projects in \toolname. Repositories vary widely in both size and difficulty. The four largest repositories (ArkLib, capless-lean, clean, lean-mlir) contribute 238 of the 500 tasks and exhibit high transitive dependency counts, reflecting deep project-specific abstraction layers. Success rates range from 0\% (IntroEffects, formal-snarks) to 100\% (veil, lean-hoare, Lentil), though the latter come from small repositories with few tasks. Across most repositories, curated context yields equal or higher success rates than full-repository context, consistent with the main text findings. Notable exceptions include pcf-lean, where full-repository context outperforms curated (71.4\% vs.\ 57.1\%), suggesting that ambient repository structure can occasionally provide useful proof cues.

\begin{table}[h]
\centering
\small
\renewcommand{\arraystretch}{1.1}
\setlength{\tabcolsep}{3pt}
\begin{tabular}{l l c cc cc cc}
\specialrule{1.2pt}{2pt}{2pt}
 & & & \multicolumn{2}{c}{\textbf{Proj Deps}} & \multicolumn{2}{c}{\textbf{Lib Deps}} & \multicolumn{2}{c}{\textbf{Success (\%)}} \\
\cmidrule(lr){4-5} \cmidrule(lr){6-7} \cmidrule(lr){8-9}
\textbf{Repository} & \textbf{Description} & \textbf{N} & \textbf{Dir.} & \textbf{Tr.} & \textbf{Dir.} & \textbf{Tr.} & \textbf{Cur.} & \textbf{Full} \\
\midrule
ArkLib & Verified SNARKs & 60 & 8.8 & 38.0 & 16.6 & 79.3 & 25.0 & 15.0 \\
capless-lean & Capability type system & 60 & 48.2 & 91.0 & 0.3 & 3.7 & 51.7 & 46.7 \\
clean & Verified ZK circuit DSL & 59 & 12.7 & 53.9 & 6.9 & 27.9 & 30.5 & 25.4 \\
lean-mlir & SSA compiler IR verif. & 59 & 12.5 & 51.3 & 7.6 & 33.7 & 10.2 & 6.8 \\
VCV-io & Crypto protocol verif. & 24 & 6.0 & 17.1 & 7.6 & 18.8 & 70.8 & 66.7 \\
iris-lean & Separation logic framework & 24 & 19.3 & 46.0 & 0.7 & 8.1 & 33.3 & 20.8 \\
loom & Verifier generation & 24 & 8.5 & 23.5 & 5.2 & 26.8 & 37.5 & 25.0 \\
splean & Separation logic proofs & 24 & 13.3 & 33.5 & 8.1 & 25.7 & 29.2 & 20.8 \\
TTBFL & Bounded universe levels & 22 & 15.5 & 34.5 & 1.1 & 11.0 & 63.6 & 59.1 \\
FVIntmax & Blockchain protocol verif. & 21 & 8.5 & 38.6 & 7.0 & 25.5 & 19.0 & 14.3 \\
juvix-lean & Juvix compiler correct. & 20 & 33.1 & 28.4 & 0.8 & 14.4 & 65.0 & 60.0 \\
lean-formal-reas. & Formal prog.\ reasoning & 20 & 7.8 & 14.2 & 0.3 & 1.6 & 90.0 & 85.0 \\
wadray\_verif. & Fixed-point arith.\ verif. & 20 & 9.4 & 11.3 & 20.1 & 24.9 & 50.0 & 50.0 \\
LeroyCompVerif & IMP compiler correct. & 19 & 12.3 & 25.7 & 1.3 & 14.2 & 26.3 & 21.1 \\
CvxLean & Convex optimization & 9 & 5.2 & 11.3 & 9.7 & 30.3 & 55.6 & 55.6 \\
pcf-lean & PCF denotational sem. & 7 & 6.6 & 25.0 & 1.4 & 8.0 & 57.1 & 71.4 \\
LeanExprEval. & Expression evaluator & 6 & 12.3 & 16.7 & 3.3 & 26.0 & 100.0 & 66.7 \\
veil & Transition system verif. & 6 & 11.0 & 13.2 & 0.7 & 6.8 & 100.0 & 100.0 \\
verified-compiler & Toy stack compiler & 6 & 7.7 & 16.7 & 2.0 & 8.3 & 16.7 & 16.7 \\
lean-hoare & Hoare logic for While & 4 & 9.5 & 5.5 & 0.5 & 4.0 & 100.0 & 100.0 \\
IntroEffects & Algebraic effects & 3 & 4.7 & 19.0 & 0.7 & 15.0 & 0.0 & 0.0 \\
formal-snarks & SNARK soundness proofs & 2 & 17.0 & 27.5 & 78.5 & 99.0 & 0.0 & 0.0 \\
Lentil & Temporal logic (TLA) & 1 & 5.0 & 15.0 & 2.0 & 2.0 & 100.0 & 100.0 \\
\specialrule{1.2pt}{2pt}{2pt}
\end{tabular}
\caption{Repository statistics and benchmark evaluation results. \textbf{N}: number of tasks from each repository. \textbf{Description}: brief summary of the project's verification target. Columns 4--7 detail the average direct and transitive dependencies for project and library components. The final columns report the best-model success rate under two settings: Curated Context and Full Context.}
\label{tab:detailed-repo-stats}
\end{table}

Table~\ref{tab:correlation} reports Spearman rank correlations between task-level features and Gemini-3-Pro success under both context regimes. We consider dependency features at two granularities (direct references vs.\ transitive closure), as well as ground-truth proof length as a proxy for intrinsic proof complexity. All features are negatively correlated with success, but structural dependency features---nesting depth ($r = -0.356$) and total transitive count ($r = -0.337$)---are substantially stronger predictors than direct counts ($|r| < 0.19$). Proof length in lines ($r = -0.254$) is moderately correlated, confirming that longer proofs are harder but that dependency structure captures difficulty beyond proof length alone. Correlations are consistent across curated and full-repository settings.

\begin{table}[h]
\centering
\small
\setlength{\tabcolsep}{4pt}
\renewcommand{\arraystretch}{1.05}
\begin{tabular}{l cc cc}
\specialrule{1.2pt}{2pt}{2pt}
\rowcolor{gray!15}
& \multicolumn{2}{c}{\textbf{Curated}} & \multicolumn{2}{c}{\textbf{Full Repo}} \\
\cmidrule(lr){2-3} \cmidrule(lr){4-5}
\textbf{Feature} & $r$ & $p$ & $r$ & $p$ \\
\midrule
Direct project deps      & $-0.128$ & $<0.01$  & $-0.156$ & $<0.001$ \\
Direct library deps      & $-0.154$ & $<0.001$ & $-0.179$ & $<0.001$ \\
Total transitive deps    & $-0.342$ & $<0.001$ & $-0.337$ & $<0.001$ \\
Nesting depth            & $-0.338$ & $<0.001$ & $-0.356$ & $<0.001$ \\
Proof length (lines)     & $-0.236$ & $<0.001$ & $-0.254$ & $<0.001$ \\
\specialrule{1.2pt}{2pt}{2pt}
\end{tabular}
\caption{Spearman rank correlations between task features and Gemini-3-Pro success (with repair). \emph{Direct} counts refer to definitions and lemmas explicitly referenced by the theorem; \emph{total transitive} includes all project and library items reachable through the dependency graph. All correlations are negative, indicating higher values are associated with lower success rates.}
\label{tab:correlation}
\end{table}

\subsection{Task Distribution and Robustness}
\label{sec:within-repo}

The 500 tasks in \toolname\ are not uniformly distributed across repositories: the four largest projects (ArkLib, capless-lean, clean, lean-mlir) contribute 238 tasks (48\%), while nine repositories contribute fewer than 10 tasks each. This reflects natural variation in project scale and maturity as well as deliberate retention of substantial developments that instantiate the repository-scale regime. After validity and non-triviality filtering, the selected fraction averages 10.4\% across repositories (median 8.6\%), ranging from 0.2\% for lean-mlir to 26\% for LeroyCompilerVerificationCourse. The five verification domains nevertheless each contribute between 16\% and 27\% of benchmark tasks. To check that our aggregate findings are not driven by cross-repository differences, we verify the key trends under two controls.

First, we compare micro-averaged success rates (across all tasks) with macro-averaged rates (averaging per-repo success rates, giving equal weight to each repository). For Gemini-3-Pro (with repair), the micro-averaged rates are 40.4\% (curated) and 35.8\% (full), while the macro-averaged rates are 49.3\% and 45.8\%. The macro average is higher because several small repositories achieve near-perfect success, but the curated-vs-full gap is consistent under both aggregations.

Second, we compute the partial Spearman correlation between nesting depth and success after residualizing out repository identity. The correlation remains significant under both regimes (partial $r = -0.208$, $p < 10^{-5}$ for both curated and full context), confirming that nesting depth predicts difficulty within repositories, not just across them.

\subsection{Error Analysis}
\label{sec:error-analysis}

\begin{table}[ht!]
\centering
\small
\renewcommand{\arraystretch}{1.05}
\begin{tabular}{l rr rr}
\specialrule{1.2pt}{2pt}{2pt}
\rowcolor{gray!15}
& \multicolumn{2}{c}{\textbf{Curated}} & \multicolumn{2}{c}{\textbf{Full Repo}} \\
\cmidrule(lr){2-3} \cmidrule(lr){4-5}
\textbf{Error Category} & \textbf{N} & \textbf{\%} & \textbf{N} & \textbf{\%} \\
\midrule
Unknown identifier      & 146 & 49.3 & 156 & 48.9 \\
Type mismatch           &  45 & 15.2 &  35 & 11.0 \\
Unsolved goals          &  32 & 10.8 &  45 & 14.1 \\
Tactic failure          &  31 & 10.5 &  37 & 11.6 \\
Other                   &  42 & 14.2 &  46 & 14.4 \\
\midrule
\textbf{Total failed}   & \textbf{296} & & \textbf{319} & \\
\specialrule{1.2pt}{2pt}{2pt}
\end{tabular}
\caption{Distribution of error types for failed Gemini-3-Pro proof attempts under both context regimes.}
\label{tab:error-analysis}
\end{table}

Table~\ref{tab:error-analysis} categorizes the Lean compiler errors produced by failed Gemini-3-Pro proof attempts (with repair) under both context regimes. Errors are classified by keyword matching on the first compiler diagnostic of each failed theorem: \emph{unknown identifier} corresponds to Lean reporting \texttt{unknown identifier} or \texttt{unknown constant}; \emph{type mismatch} to \texttt{type mismatch} or \texttt{application type mismatch}; \emph{unsolved goals} to goals remaining after the proof script terminates; and \emph{tactic failure} to any tactic-specific error (e.g., \texttt{tactic `rewrite' failed}). Each theorem is deduplicated across runs, and only theorems where no attempt succeeded are counted.

The dominant failure mode under both regimes is \emph{unknown identifier} ($\sim$49\%), indicating that the model frequently hallucinates references to definitions or lemmas that do not exist in the project. This suggests that models struggle to accurately recall project-specific names from the provided context. The error distribution is similar across curated and full-repository settings, with a slight increase in unsolved goals under full context (14.1\% vs.\ 10.8\%), possibly reflecting cases where the model constructs a partially correct proof but cannot close all subgoals when distracted by irrelevant repository content.

\end{document}